\documentclass[journal]{IEEEtran}
\usepackage{blindtext,graphicx,booktabs,amsmath}
\usepackage{amsmath}
\usepackage{multirow}
\usepackage{hhline}
\usepackage[english]{babel}
\usepackage[utf8]{inputenc}
\usepackage{algorithm}
\usepackage{multirow,tabularx}
\usepackage[noend]{algpseudocode}

\begin{document}
\title{Interpenetrating Cooperative Localization in Dynamic Connected Vehicle Networks
% : Mathematical Analysis and Real World Experiments
}
%Rao-Blackwellized Partical Filters and DSRC Performance Based on Safety Pilot Model}

%on Vehicle Networks with Real-World Traffic} %Optimized Vehicle Networks}
%Cooperative Road Configuration}

\author{Huajing Zhao$^{1}$, Zhaobin Mo$^{2}$, Macheng Shen$^{3}$, Jing Sun$^{4}$, Ding Zhao$^{1}$

\thanks{*This work is funded by the Mobility Transformation Center at the
University of Michigan with grant No. N021548.}
\thanks{$^{1}$H. Zhao and D. Zhao are with the Department of Mechanical Engineering, University of Michigan, Ann Arbor,
 Ann Arbor, MI, 48109. (corresponding author: Ding Zhao zhaoding@umich.edu)}
\thanks{$^{2}$Z. Mo is with the Mechanical Engineering at the Tsinghua University, Beijing, China, 100084.}%
\thanks{$^{3}$M. Shen is with the Mechanical Engineering at the Massachusetts Institute of Technology, Cambridge, MA, USA, 02139.}%
\thanks{$^{4}$J. Sun is with the Department of Naval Architecture and Marine Engineering, University of Michigan, Ann Arbor,
MI, 48109.}

% <-this % stops a space
}

% \markboth{IEEE TRANSACTIONS ON INTELLIGENT TRANSPORTATION SYSTEMS,~Vol.~XX, No.~XX, December~2017}%
% {Shell \MakeLowercase{\textit{et al.}}: Bare Demo of IEEEtran.cls for Journals}

\maketitle

\begin{abstract}
In this paper, we proposed the Interpenetrating Cooperative Localization (ICL) method to enhance the localization accuracy in dynamic connected vehicle networks. This mechanism makes the information from one group of connected vehicles interpenetrate to other groups without full communication between all nodes, thus improving the utility of information in a low connected vehicle penetration situation. We tested the approach using the
dynamic traffic data collected in the Safety Pilot Model Deployment program in Ann Arbor Michigan, USA, with dynamic changing networks due to the traveling of vehicles and packet drops of the Dedicated Short-Range Communication. Results show enhancement of localization accuracy with errors reduced by up to 70\% even in complex dynamic scenarios.
\end{abstract}
% \begin{IEEEkeywords}
% IEEEtran, journal, \LaTeX, paper, template.
% \end{IEEEkeywords}
\IEEEpeerreviewmaketitle
\section{Introduction}

With emerging automation in the vehicle domain,
approaches that enhance the localization accuracy for automobiles
are subjected to increasing new applications \cite{zhao2017accelerated}. In general, a Global Navigation Satellite System
(GNSS) receiver calculates a vehicle's position from pseudo-range
measurements of multiple satellites. During this process, pseudo-ranges
error results in a position displacement of several meters. Pseudo-ranges
error can be decomposed into two components: common error, which is due to
satellite clock error, ionospheric and tropospheric delays; and
non-common error, which is due to receiver noise, receiver clock error
and multipath error
 \cite{shen2017improving}. 
%, to achieve the objectives
%of collision avoidance, lane-changing assistance, danger warning, etc.
While non-common errors are hard to eliminate, modern technologies have provided multiple ways to 
%improve GNSS accuracy by reducing common-error. For example, Differential GNSS (DGNSS) uses a network of fixed reference stations to eliminate common biases, which improves the localization accuracy to the sub-meter
level. Real-Time Kinematic (RTK) \cite{wang1999stochastic}
technique enhances
accuracy to centimeter-level by using carrier phase measurements to provide
real-time corrections\cite{shen2017improving}. However, most of these techniques require additional devices or being expensive, thus not economical enough for the localization tasks of general automobiles. 

\begin{figure}[ht]
\includegraphics[width=\linewidth]{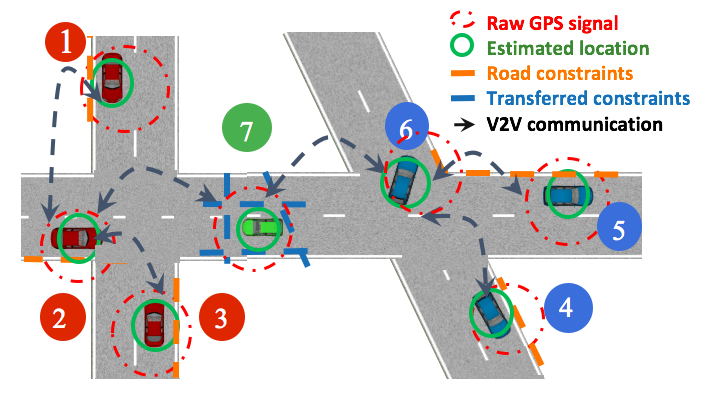}
%ppt.png}
%4.png}
%{fig/CMM_cover_img.jpg}
     \caption{Illusion for Interpenetrating Cooperative Localization using road constrains in urban traffic.
    %Red circles represent GNSS raw measurement, green circles represent CMM measurement, yellow dash lines represent road constrains used. 
    The network is not fully connected, but interpenetrating: each car has communication to only some of its neighbors to gain CMM information. Here, the vehicles with number 1, 2, 3 are connected in one group, while vehicles with 4, 5, 6 are in another. The RBPF measurement are fused within group and transfer to vehicle 7 (the host) by 2 and 6.}
     \label{fig:coverImg}
 \end{figure}

Recently, connected vehicle technologies are drawing increasing attention from vehicular industrial. For example, 
Ford already has 700,000 connected cars on the road, and have made the partnership with Qualcomm to develop connected vehicle technology to ease congestion, planning to have every new Ford car to be connected by 2019 \cite{ford2018news}. Connected vehicle technologies provide an alternate to achieve high accuracy with low-cost by using Cooperative Map Match (CMM) \cite{rohani2016novel,obst2012cooperative,rife,IEEE_magazine}, which achieves accuracy up to lane-level accuracy and available for real-life implementation. Fig. \ref{fig:coverImg} shows the raw GNSS positioning (red dot and red ellipsoid) and the corrected one by CMM (green ellipsoid), and the effects of road constraints are demonstrated.

% CMM reduces the common errors in vehicular positioning by matching the raw GNSS measured locations for a group
% of connected vehicles to a digital
% map. 
%GPS-enabled smartphones are typically accurate to within a 4.9 m (16 ft), but the performance worsens near buildings and trees. 
Our previous work on CMM \cite{shen2017improving,DBLP:journals/corr/ShenZS17a} have investigated the theories and methodologies for using Rao-Blackwellized Particle Filter (RBPF) to improve the localization accuracy, which was achieved by using road constrains of a group of vehicles under the same set of satellites' measurement to remove common-errors.
%The government commits to broadcasting the GPS signal in space with a global average user range error (URE) of $\le$ 7.8 m (25.6 ft.) with 95$\%$ probability [2].
%On May 11, 2016, the global average URE was ≤0.715 m (2.3 ft.), 95$\%$ of the time.
%
However, the following issues
the communication distance of DSRC is limited, and vehicles far away from each other cannot directly communicate. Secondly, the vehicle configuration and CMM algorithm needs
%there are normally far more cars within a communication range of the connected vehicles, and 
% the impact analysis with 
%test with real data experiments based on the GPS and
communication devices. Thirdly, the CMM algorithm also needs evaluation in dynamic situation,
%trajectories of vehicle members
and should have the packet losses during communication considered. This paper thus addresses the above-mentioned problems, proposed a interpenetrating cooperative localization mechanism in a decentralized fusion manner, and provided an implementation under dynamic real-world traffic data abstracted from Safety-Pilot Model Deployment (SPMD) database.
% The DSRC packet drop is also applied to better simulate the real-world communications.
%
% The nominal accuracy of pseudo-ranges
% for a single-band receiver is about 10 to 20 meters, which
% results in a position error of several meters \cite{zhao2017accelerated}. Without further
% improvement, this crude GNSS error is too large for many
% safety functions as the lane of vehicles cannot be robust.
Comparing to previous publications where every car stays in the same position and has constant communications to the connected neighbors,
%it was connected to from the beginning, 
this work emphasized more on practical implementation.
In this work, we proposed the Interpenetrating Cooperative Localization (ICL) method, which is a distributed fusion mechanism within the vehicular ad-hoc network. The implementation is then presented on real-world traffic data, which is based on the historical on-road vehicle records traveling simultaneously in Ann Arbor city. This mechanism makes the information from each node reachable within the network by other nodes without direct communication, thus improving the overall localization accuracy and robustness. The dynamic model, decentralized cooperative map matching and practical packet losses in the connected vehicle network
regarding DSRC are included.  

% provide a improve model with dynamic scenario, and implemented on an optimized vehicle network structure based on the real-world road condition database. 
 
% To select the optimal group of vehicles to form network configuration, a CE method is implemented, together with

In the following sections, we presented the mechanism and implementation
of both traditional and interpenetrating CMM on real-world traffic scenarios. 
%In Section II, related works about GNSS localization, CMM and DSRC are introduced.
In Section II, we introduce the motivations on developing  interpenetrating vehicle network localization using decentralized CMM, and
the mechanism is presented under dynamic assumption. In Section V, we introduced the mechanism for Interpenetrating Cooperative Localization, and the procedure involved using RBPF. In Section IV, we present the formulation of decentralized vehicle network, the data used for vehicle network simulation, and the DSRC related packet losses. In Section V, we demonstrate the model formulations for simulating real-world traffics. In Section VI, simulation
results are presented under stationary and dynamic scenarios using real-world vehicle data,
with simulated satellite measurement models using Satellite Navigation
TOOLBOX 3.0.
%, followed by the performance analysis of the RBPF. 
Discussions and Conclusions are presented in Section
VII and Section VIII.
%\section{Related Works}
% Vehicle related applications in most real-life traffic scenarios, such as driver safety assisting system and 
% %Automatic Guided System in 
% self-driving cars, always require a relatively
% high level of accuracy in vehicular position information.
%, so as to achieve the objectives
%of collision avoidance, lane-changing assistance, danger warning, etc.

%For these reasons some innovative approaches, combined with mobile communication technologies in vehicular networks, have been developed to enhance positioning accuracy \cite{Bagloee2016,suman2000vehicle}. 

\section{Interpenetrating Cooperative Localization}

%\section{Motivation}
\indent The capability of CMM to
mitigate the biases from the localization results obtained
from low-cost GNSS receivers alone has been investigated by researchers such as Rohani et al. \cite{rohani2016novel}. Our prior work \cite{shen2017improving} has developed an RBPF that filters the uncorrelated error and eliminates the effects of correlated error by cooperative map matching (CMM). CMM, first proposed in Rohani et al. \cite{rohani2016novel}, is a method that matches the GNSS positioning of a group of vehicles to a digital map and corrects the biased positioning by enforcing the road constraints. 
%  The effectiveness of CMM depends on the diversity of the road constraints.
In prior work \cite{shen2017impact}, we quantified the correlation between the diversity of the road constraints and the CMM localization error, and developed algorithms that optimally design the connection network subject to limited communication bandwidth \cite{shen2017optimization}. 
\subsection{Why Decentralized Structure?}
\indent The methods developed \cite{shen2017impact,shen2017improving} implicitly assumed a centralized architecture that gathers all the GNSS measurements for fusion and optimization. In real vehicular networks, especially the large ones, a decentralized architecture is more desirable and realistic for the following reasons:
\begin{enumerate}
 \item The communication distance of DSRC is limited. Vehicles far away from each other cannot directly communicate.
 \item The communication bandwidth is limited. Each vehicle can only communicate with up to 30 other vehicles without significant package loss \cite{ramachandran2007experimental}.
 \item The on-board computational capacity is limited. As the computational complexity grows with the number of vehicles involved, on-line filtering of raw data could become challenging. 
 \end{enumerate}

\indent While increasing the number of participating vehicles in the network can further localization errors, the limits of communication range and bandwidth make the configuration of vehicle membership crucial for real-life vehicle network optimization \cite{DBLP:journals/corr/ShenZS17a,shen2017impact}. One straightforward way to overcome the communication constraints is to implement individual RBPF on each vehicle using the locally available GNSS measurements from itself and all of its neighbors, to update the estimation. Unlike centralized CMM, decentralized CMM networks require less computation capacity, and is thus a more realistic structure for real-life connective vehicular networks. It uses a distributed
fusion mechanism within the vehicular ad-hoc network, which makes the information from each node
within the network accessible by other nodes without direct communication. In this way, the required communication density is reduced, while the richness of information is preserved.

\subsection{Interpenetrating Vehicular Networks}
It is expected that the reduced number of vehicles involved in commutation would lead to a sparse network where the number of links connecting nodes is sparse. This could, in turn, lead to large localization error, as there would not be enough road constraints to mitigate the common GNSS error. Given this limitation, decentralized CMM along with Interpenetrating Cooperation Networks is a desired solution for balancing the communication capacity and the localization accuracy. The pseudo-code for this procedure is presented in Algorithm. \ref{alg:semi-intp}

% 接受信号 -》 判断掉包 -》选车 -》CMM -》发出去和其他人fuse

\begin{algorithm}
\caption{Interpenetrating CMM}\label{alg:semi-intp}
\begin{algorithmic}[1]
\Procedure{Interpenetrating Network}
{GNSS Raw Measurement
}
%\Comment{The g.c.d. of a and b}
\State Receive Packets from Neighbors
\State Determine whether there is a packet loss
\State Choose available neighboring vehicles to connect
\State Update RBPF estimations with measurements from neighbors
\State Fuse the RBPF estimations with connected vehicles
\State \textbf{return} CMM {Enhanced Measurement}
%\Comment{The gcd is b}
\EndProcedure
\end{algorithmic}
\end{algorithm}

In interpenetrating vehicle networks, each vehicle not only uses the measurements from neighbors to update their own RBPF estimations, but also fuse the RBPF estimations of neighbors. This provides an innovative method to partially information from nodes that are not directly connected to the receiver without expanding communication capacity. As the estimations are represented by RBPFs, the fusion can be performed simply stacking the particles and eliminating those inconsistent with the road constraints. Information from any node can be propagated to any other node by repeated local fusion as long as the graph representing the network is connected. We also expect that the resulted average localization error would be much smaller than that of the decentralized CMM without fusion. Furthermore, the localization error can be minimized if the fusion is designed to optimize certain error criterion.

\subsection{CMM for Dynamic Vehicle Network}
Cooperative map matching improves
navigation solutions by to correct the common
localization error. Assuming that most vehicles travel
within lanes, the correction to the common localization
biases can be determined so that the corrected positions
of a group of vehicles best fit the road map. The pseudo code of the proposed RBPF is
shown in Algorithm.\ref{alg:rbpf}.
%This requires communication within the group of vehicles, hereafter referred to as cooperative map matching (CMM). 

\begin{algorithm}
\caption{RBPF for CMM vehicle localization}\label{alg:rbpf}
\begin{algorithmic}[1]
\Procedure{RBPF}{$C_{t-1}^k,X_{t-1}^k,Z_{t-1}^k$}
%\Comment{The g.c.d. of a and b}
\State Predict $C_t$ and $X_t$ for vehicle i = 1 : $N_v$ %according to Eq. (3,5) 
% $r\gets a\bmod b$
%\While{$r\not=0$}
%\Comment{We have the answer if r is 0}
\State Determine the indicator variable
\State Calculate particle weights and update $X_t$
\State Modify particle weights
% \EndWhile\label{euclidendwhile}
\State Resample particles
\State \textbf{return} $C_{t}^k,X_{t}^k,\Sigma_{t}^k$
%\Comment{The gcd is b}
\EndProcedure
\end{algorithmic}
\end{algorithm}
 % to reduce common error from observed positions using CMM.
The vehicle cooperative localization has been formulated in this work as a Bayesian filtering problem which estimates the joint posterior distribution of the GNSS common biases and the states of the vehicles conditioned on the raw pseudo-range measurements. First, we introduce the expression for using particle filter to find vehicle states in a dynamic network. For each vehicle $k$ at timestep $t$, the GNSS pseudo-ranges, pseudo-ranges common bias, pseudo-ranges covariance, and vehicle state are denoted as $Z_t^k$, $C_t^k$, $\Sigma_t^k$, and $X_{t}^k$, respectively. Given the assumption that the non-common errors of different vehicles are uncorrelated, it can be implied that conditioned on the common biases, the posterior distributions of the vehicle states are independent of each other, which means that the joint distribution can be factorized as in Eqn. \ref{1}.

\begin{equation}
\begin{aligned}
p(C_{1:t}^{1:N_{s}},X_{1:t}^{1:N_v}|Z_{1:t}) = p(X_{1:t}^{1:N_v}|C_{1:t}^{1:N_{s}},Z_{1:t}) p(C_{1:t}^{1:N_{s}}|Z_{1:t}) 
\\
= \prod_{i=1}^{N_v} p(X_{1:t}^{i}|C_{1:t}^{1:N_s},Z_{1:t}) p(C_{1:t}^{1:N_s}|Z_{1:t})
\label{1}
\end{aligned}
\end{equation}
where $N_{s}$ and $N_v$ are the number of satellites and vehicles
respectively, and the superscript 1:$N_s$ and 1:$N_v$ are
shorthands for the corresponding variables of all the satellites
and all the vehicles. The subscript 1:t is the shorthand for
the corresponding variables of all the time instances.

The RBPF exploits this conditional independence property
for efficient inference of the pseudo-range common biases and
the vehicle states given the pseudo-range observations. The
posterior distribution of the common biases $p(C_{1:t}^{1:N_s}
 |Z_{1:t})$ is
estimated by particle filter, and the distributions of the vehicle
states conditioned on the common biases $p(X_{1:t}^{i}|C_{1:t}^{1:N_s},Z_{1:t})$
are independent of each other and estimated by a set of
EKFs whose dimension is the dimension of the state vector
\cite{shen2017improving}. 
%\subsection{Prediction of State and Update of Measurement}
The RBPF uses a predict-update framework to estimate the common error. In the prediction step, vehicle positions with common error are predicted by each particle with random Gaussian noise added. 
In the update step, those particles incompatible with the road constraints have a large probability of being eliminated. 
%As a result, the RBPF estimation after map matching is biased towards the direction where the particles obey the road constraints.
With the assumption that common biases vary slowly over time, we model the time variation of the
common biases as a first-order Gaussian-Markov process. 

Neglecting the multipath error, the GNSS predicted mean of pseudo-range measurement model between satellite j
and vehicle i is 
\begin{equation}
\begin{aligned}
Z_t^{i,j} = \|p_t^i - s_t^j\| + C_t^j + b_t^j
\label{3}
\end{aligned}
\end{equation}
where $p_i^t$ is the position of the vehicle and $s_j^t$ is the satellite
position.
% \textbf{To be edited} Rohani et al. [5] presented 
The
CMM algorithm then use a particle-based method that involves
only the vehicle positions’ estimation of the current epoch from vehicle members.
%to address these three difficulties 
The recursive prediction-update equations are presented
]as follow:

\begin{equation}
\begin{aligned}
C_t^j = C_{t-1}^j + w_t^j \Delta t
\label{2}
\end{aligned}
\end{equation}
where $w_t^j \sim N(0,\sigma_x^2)$, $\sigma_c^2$ representing the variance of common bias drift, $\Delta t$ is length of the time interval between two successive updates of the states and
$j = \{1, 2, ...,N_{s}\}$ is the index for satellites.
% As we assume that 
% only the horizontal positions and velocities need be modeled explicitly, t
The dynamic state vector of the $i$th vehicle can be modeled by the following expression:

\begin{equation}
\begin{aligned}
X_i^t = [x_i^t \quad \dot{x}_i^t \quad y_i^t \quad \dot{y}_i^t \quad b_i^t \quad \dot{b}_i^t]
\label{2}
\end{aligned}
\end{equation}
where $x_i^t$ and $ y_i^t$ are the horizontal positions; $\dot{x}_i^t$ and $\dot{y}_t^i$ are the horizontal velocities; and $b_i^t$ and $\dot{b}_i^t$ are the receiver clock bias and drift.
The mean state and covariance for $i$th vehicle at time $t$ are then propagated by 
$\bar{X}_t^i = AX_{t-1}^i$, and $\bar{\Sigma}_t^i = AX_{t-1}^iA^T + R_t$, where 
\begin{equation}
  A = \left[
\begin{matrix}
    1 & \Delta t & 0 & 0  & 0 & 0\\
    0 & 1 & 0 & 0  & 0 & 0\\
    0 & 0 & 1 & \Delta t  & 0 & 0\\
    0 & 0 & 0 & 1  & 0 & 0\\
    0 & 0 & 0 & 0  & 1 & \Delta t\\
    0 & 0 & 0 & 0  & 0 & 1\\
\end{matrix}
\right],
\label{2}
\end{equation}
and $\bar{X}_t^i$ is the predicted mean, $\bar{\Sigma}_t^i$ is the covariance matrix of the state vector, $R_t$ considers the variances of the horizontal acceleration, clock bias and drift time derivatives. 

The effect of non-common pseudo-range error was
considered by a weighted road map approach to preserve
consistency, and by tracking
the origin of the common bias corrections from different
vehicles and fusing only those corrections from independent
sources to avoid data incest, over-convergence can be effectively avoided.
%though some of the correlated corrections containing additional information have to be discarded. 
%\textbf{RBPF: either more in detail or get simpler}
The weights of the particles are calculated accordingly. Each vehicle within the network then take particles
from their neighboring vehicles and stack them together
with its own particles. The CMM algorithm then calculate weights according to the map matching and
resample to downsize the number of the particles to
the original size, as demonstrated in Algorithm. \ref{alg:rbpf}. The RBPF estimated common bias $C_{t}^k$, vehicle state $X_t^k$, and covariance $\Sigma_t^k$ are then returned as outputs for each vehicle $k$ at timestep $t$, which are calculated by Monte Carlo Integration. The propagation follows the rule of particle filters, the detailed expression refers to our previous work \cite{DBLP:journals/corr/ShenZS17a} in its Section. 3. 

% 1) Predict Ct and Xt according to Eq. (3,5)
% for vehicle i = 1 : Nv

% 2) Determine the indicator variable 

% 3) Calculate particle weights and update Xt

% 4) Modify particle weights

% 5) Resample
% end RBPF

\section{Formulation of Decentralized Network}

Before implementing decentralized CMM to produce sparse network with interpenetrating message passing,
we first introduce the dynamic vehicle model formulation from SPMD database, and study how to combine the optimization of vehicle connection with RBPF under dynamic situations.

\subsection{Vehicle Selection from Real-world Database}

To simulate the real-world traffic, we implement the historical vehicle driving data abstracted from the Safety Pilot database, which involves installing devices in about 2800 vehicles, including cars, trucks, and buses \cite{sottile2011hybrid}.
%A Rao-Blackwellized particle filter (RBPF) is then used to jointly estimate the common biases of the pseudo-ranges and the vehicle positions \cite{zhao2017accelerated}.
%The quantitative relationship between the estimation error and the road constraints which has also been systematically investigated to provide insights.
In this paper, we utilize a small data set of the Safety Pilot database for simulation purpose, which was collected from 136 vehicles running for about 62 months and contains about 100,000 trajectory record. 
\begin{figure}[ht]
\includegraphics[width=\linewidth]{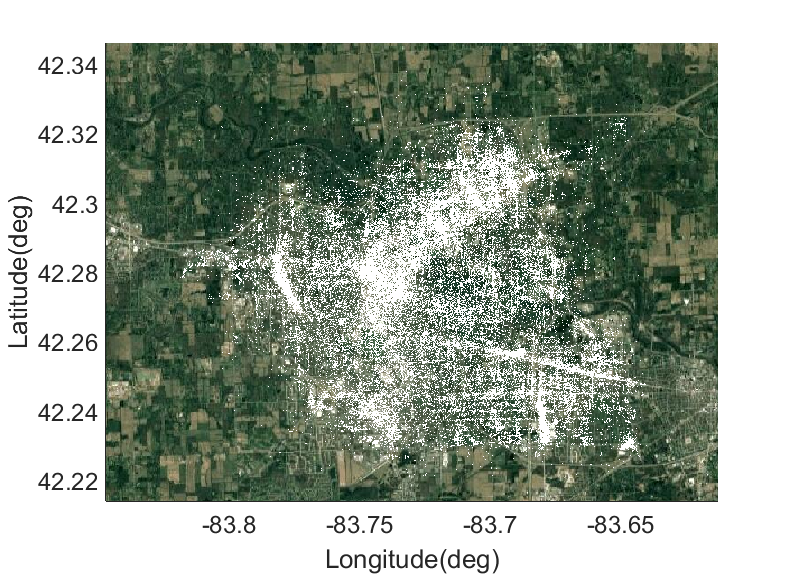}
     \caption{Distribution of the 100,000 trajectory records. The white dot indicates the central points of the minimal rectangular that can encompass the trajectory }
     \label{fig:100kcentralpoints}
 \end{figure}
 
Fig.\ref{fig:100kcentralpoints} indicates the distribution of the trajectories, in which the central points of the minimal rectangular that can encompass the trajectory is used to include the massive trajectories in one image. The trajectory information we extracted include latitude, longitude, speed and heading angle of the vehicles at each 0.1 second time interval. The selection range is constricted to an urban area with the range of latitude and longitude to be (-83.82,-83.64) and (42.22,42.34) respectively.

\begin{figure}[ht]
\includegraphics[width=\linewidth]{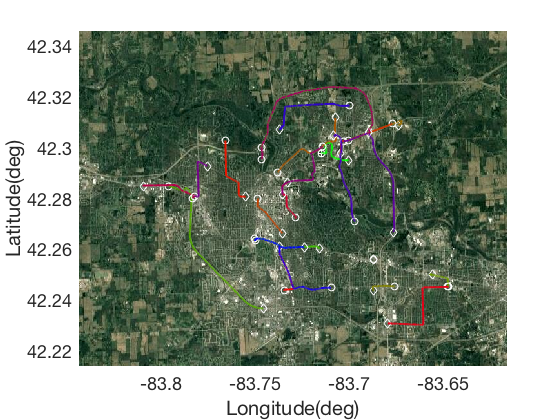}
     \caption{Vehicle trajectories  selected in a 24 vehicle connected network}
     \label{fig:24vehiDistb}
 \end{figure}

\begin{figure}[ht]
\includegraphics[width=\linewidth]{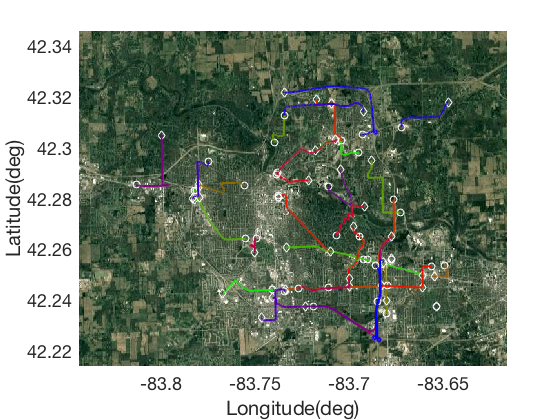}
     \caption{Vehicle trajectories selected in a 38 vehicle connected network}
     \label{fig:38vehiDistb}
 \end{figure}
 
 \begin{figure}[ht]
\includegraphics[width=\linewidth]{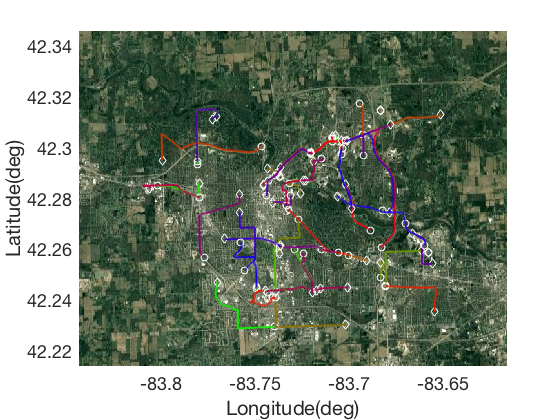}
     \caption{Vehicle trajectories  selected in a 50 vehicle connected network}
     \label{fig:50vehiDistb}
 \end{figure}

As the simulation time in this paper is set to be 300 seconds, we firstly select the trajectories longer than 300 seconds and then extract the 300-seconds subtrajectory from them. What’s more, to maintain the continuity of the trajectory, we eliminate the trajectory which contains the sample point where the heading angle changes more than 10 degrees in 0.1 seconds. To keep the original trajectory distribution, trajectories are randomly selected from the 100,000 records. This gives the trajectories in area with tense traffic flows to have higher probability to be chosen.

In this way, 24,38,50 vehicle networks are selected to further simulate different density of the network, each shown in Fig. \ref{fig:24vehiDistb}, Fig. \ref{fig:38vehiDistb} and Fig. \ref{fig:50vehiDistb}. The circle and diamond indicate the starting and ending point of the trajectory respectively.

%To simulate the real-world traffic, we implement the historical vehicle driving data abstracted from the Safety Pilot database, which simulate based on the past 3 years driving data of Ann Arbor city to provide the most representative corresponding routes for 24, 38, 50 vehicle networks according to actual road conditions, each shown in Fig. \ref{fig:24vehiDistb}, Fig. \ref{fig:38vehiDistb} and Fig. \ref{fig:50vehiDistb}.  Trajectories are randomly selected from the 100k record trajectories from the database, based on the central point of the smallest rectangle which can encompass the trajectory.
%This provides a method which gave the trajectories in an area with tense traffic flows to have a higher probability to be chosen. A localization method was proposed based on the optimized configuration of vehicle connections.

The main draw back of the method is that the trajectories are selected from the record of last three years, which means the network may not exist ever before even though each trajectory is historic data. However, as we select the networks from a large data set (100,000 records), they are representative and is valid enough at simulation stage.

A localization method was then proposed based on the optimized configuration of vehicle connections.

\subsection{Modeling RBPF dynamics and fusion}
In this section, we introduce the notations used to describe the vehicular network, the mathematical modeling of the RBPF dynamics and the decentralized optimization of the fusion mechanism. 

We define the connected vehicle network as an undirected graph $(V,E)\in G$, where $V=\{1,2,3,...N\}$ is the set of nodes each represents a vehicle within the network and $E$ is the set of edges within the network. If two vehicles $i$ and $j$ are within communication ranges, then there is an edge $(i,j)\in E$ between these two nodes. Nonetheless, it does not imply that there has to be a communication between these two vehicles. We use $x_i$ to denote the common error estimated by vehicle $i$, which is calculated by averaging all the common error estimations, represented by the particles belonging to vehicle $i$.
\indent The divergence in the common error estimation can be mitigated by richer information of road constraints through repeated fusions of the estimated common error from neighboring nodes. 
As both the measured position and the predicted position of each vehicle would be passed to its neighboring node,
hypothetically, if the fusion is conducted N times, where N is equal to or larger than the diameter of the network
(the largest path length between any two nodes), then each node implicitly utilizes the road constraints "seen" of all the nodes of the network, as these constrains are embedded in the RBPFs of the nodes.\\
% I tried my best to express this; does it make sense?
% "As both the measured position and the estimated position of $i$ after fusing information from all $j$s would be passed to the next node, and thus the estimated common error incorporates the information about the road constraints"
\indent We, therefore, propose a fusion mechanism for the CMM vehicular network using RBPF described as follows:
\begin{enumerate}
 \item For each vehicle within the network, take particles from their neighboring vehicles and stack them together with its own particles.
 \item Calculate weights according to the map matching and resample to downsize the number of the particles to the original size.
 \item Obtain the numbers of particles taken from neighboring vehicles by optimizing certain criterion.
 \end{enumerate}
\indent We model the evolution of the networked RBPF estimations as a coupled linear system, described by the following equation:
\begin{equation}
x_i(t+1)=\sum\limits_{j=1}^N a_{i,j}(t)x_j(t)+w_i(t),
\label{1}
\end{equation}
where
\begin{equation}
a_{i,j}(t)=0,\forall (i,j)\notin E.
\end{equation}
\indent Here, $x_i(t)$ denotes the estimated common error of vehicle $i$ at timestep $t$, while $w_i(t)$ denotes a random Gaussian noise applied to $i$ at timestep $t$. %%%%%
Eq. \ref{1} means that the estimated common error after fusion equals to a weighted average over the neighboring estimation and the own estimation at the previous time instance, added by some random noise from the RBPF prediction. This equation can explain the reason why the RBPF diverges if none of the vehicles fuses estimations from other vehicles, simply by taking $a_{i,j}=0,i\neq j; a_{i,i}=1$. This represents a discrete random walk that is driven by the random force $w_i(t)$ and can be unbounded. We expect that this linear modeling of the particle filter dynamics is an approximation to the true dynamics when all the particles are reasonably close to the true common error such that all the particles receive weights of comparable magnitude. As a result, the resampling process described in the fusion mechanism is like a noisy weighted averaging over the particle estimations. 
\subsection{Minimize variance}
\indent Eq. \ref{1} leaves freedom for us to select the fusion coefficients $a_{i,j}$ by selecting the different numbers of particles taken from neighboring vehicles. It is desirable to select the numbers of particles such that the averaged CMM localization error over the whole network is minimized. That is, we want to choose $a_{i,j}$ subject to the communication constraints such that we can minimize 
\begin{equation}
\begin{aligned}
J&=\frac{1}{N}\sum\limits_{i=1}^N (x_i(t+1)-c(t+1))^2\\
&=\frac{1}{N}\sum\limits_{i=1}^N(x_i(t+1)-\bar{x}(t+1))^2\\
&+\sum\limits_{i=1}^N(\bar{x}(t+1)-c(t+1))^2,
\end{aligned}
\label{error_decom}
\end{equation}
where $c(t+1)$ is the true common error at the time instance $t+1$, which is assumed to be slowly time-varying; $\bar{x}(t+1)=\frac{1}{N}\sum\limits_{i=1}^N x_i(t+1)$ is the mean of the common error estimation over the whole network.\\
\indent Eq. (\ref{error_decom}) has a clear interpretation: The objective $J$ represents the averaged squared estimation error over the network. It can be decomposed into two parts: The variance of the estimations over the network and the squared error of the mean estimation over the network. However, it is impossible to directly minimize the objective $J$, as the true common error is unknown. We, thereby, aim to minimize the variance in the hope that by doing so, the original objective functions $J$ remains small. We present the following arguments for this choice:
\begin{itemize}
\item The original objective $J$ is always larger of equal to the variance. It is necessary to keep the variance small to keep $J$ small.
\item The variance can be calculated given all the estimations within the network in a decentralized manner, with $\bar{x}$ obtained by distributed consensus algorithms.
\item As the variance becomes small, all the estimations within the network tend to the same estimation, which could lead to a small estimation error of the mean as a result of CMM. The original objective $J$ would also be small.
\end{itemize}
\indent Given the aforementioned rationals, we formulate the following optimization problem:
\begin{equation}
\begin{aligned}
minimize\mbox{}&\tilde{J}=\frac{1}{N}\sum\limits_{i=1}^N(x_i(t+1)-\bar{x}(t+1))^2,\\
subject\mbox{ }to\mbox{ }&x_i(t+1)=\sum\limits_{j=1}^N a_{i,j}(t)x_j(t),\\
&a_{i,j}(t)=0,\forall (i,j)\notin E,\\
and\mbox{ }& 0\leq a_{i,j}(t)\leq 1.
\end{aligned}
\label{opt_problem}
\end{equation}
\indent The fusion coefficients $a_{i,j}(t)$ are the free variables to be selected. Note that the objective function is the sum of all local objective functions, Eq. (\ref{opt_problem}) is essentially a decentralized quadratic programming.

\subsection{Accelerate Consensus}
In Section III.C, we show that it is necessary to bound the variance in order to bound the mean squared estimation error. The intuition is that as the disagreement between the nodal estimations becomes small, the estimation error tend to be bounded by map matching. Besides variance, the convergence rate of Eq. (\ref{1}) is another metric that measures the agreement and disagreement of node decisions within a network. It is expected that by maximizing the convergence rate, the mean squared localization error would also be small. It has been well studied how to maximize the convergence rate given a network with fixed structure in Xiao \cite{xiao2004fast}. We briefly mention the major results here:\\
Given a linear consensus iteration:
\begin{equation}    
X(t+1)=PX(t),
\end{equation}
where
\begin{equation}
P_{i,j}=0,\forall (i,j)\notin E.
\end{equation}
The asymptotic convergence rate is defined as:
\begin{equation}
r_{asym}(P)=\mathop{sup}\limits_{X(0)\neq \bar{X}}\lim_{t \to \infty}(\frac{||X(t)-\bar{X}||_2}{||X(0)-\bar{X}||_2})^{\frac{1}{t}},
\end{equation}
where $\bar{X}$ is the fixed point of this linear iteration.\\
\indent There is a systematic approach to maximize $r_{asym}$ by solving an optimization problem that requires the knowledge of the network topology. This would be problematic for all problem because of the communication limit. An alternative approach that does not give optimal convergence rate but only uses local topology is given below \cite{xiao2004fast}:
\begin{equation}
P_{i,j}=\frac{1}{max\{d_i,d_j\}},
\label{max_degree}
\end{equation}
where $d_i$ is the degree of the node $i$, which is equal to the number of neighbors of the node $i$. This approach guarantees convergence as well as reasonably fast convergence rate.

\section{Formulation of Dedicated Short Range Communication}

Dedicated Short Range Communication (DSRC) is a vital topic in practical connected vehicle domain.  
%The U.S. Federal Communications Commission has
 %allocated 75MHz of spectrum band around 5.9 GHz for
DSRC allows both vehicle-to-vehicle (V2V) and vehicle-to-infrastructure (V2I)
communications, which provide support for vehicular networking applications. Vehicular telematics researches include standardization efforts
such as VSCC in North America, V2V consortium in Europe, IEEE 802.11p working
group, and some research projects such as CVIS, Fleetnet and SafeSpot \cite{schoch2008communication}. In the real world, the quality of vehicle information transmission decreases with the increase in packet loss and delay
in vehicular wireless communications during the communication and positioning process, which can be simulated according to DSRC. 
%Recently, DSRC - based V2V technique has been developed and validated in the real world. 
% As is proved in experiments [7],. 
% Timely warning messages transmitted by a braking or
% slowly moving vehicle enable approaching traffic to take appropriate action such as slowing
% down or changing lanes earlier than they could have without the warning messages, and thereby
% can reduce the chance of crashes or chain collisions.
% The U.S. Federal Communications Commission has
% allocated 75MHz of spectrum band around 5.9 GHz for
Here we consider the performance of the V2V
communication system in the presence of packet loss. We assume that all vehicles in the network are equipped with the 802.11p communication unit.
% Safety-related applications are geared primarily toward avoiding the risk of car accidents,
% by using cooperative collision warning, precrash sensing, or lane change and traffic violation warnings. 
%These applications all have real-time constraints, which always rely on one-hop
%broadcasting or multihop V2V and vehicle-to-infrastructure (V2I) communications. 

\subsection{DSRC on Connected Vehicle Application}
\begin{figure}[ht]
\includegraphics[width=\linewidth]{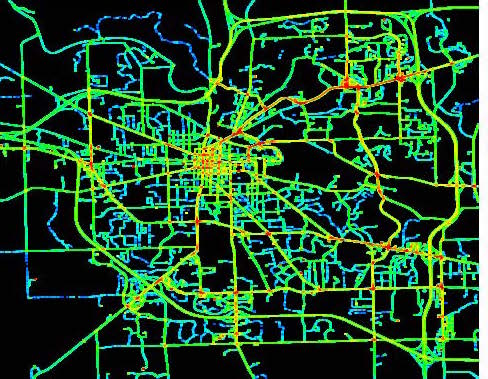}
     \caption{Vehicle density distribution on two successive weekdays; Red: dense; Green: medium; Blue: sparse}
     \label{fig:veh_density}
 \end{figure}

For the connected vehicles, the quality of vehicle safety applications degrades with the increase in packet loss and delay
in vehicular wireless communications. Possible factors that affect the communication success include hardware or software malfunctions, passing by a large vehicle, antenna location and installation, snow accumulation covering the RSE antenna and its effect on ground reflection. Fig. \ref{fig:veh_density} shown the vehicle density distribution on two successive weekdays in Ann Arbor City, which is based on the same database for the vehicle trajectories simulated for this work.

The performance of DSRC related communication range was studied in previous work \cite{DBLP:journals/corr/HuangZP16}. To characterize the performance of DSRC, the following metrics are defined.
Packet Delivery Ratio (PDR) is the ratio of successful communication events to the total number of transmission attempts at a given distance between two DSRC units.
Maximum Range (MR) is the maximum distance at which the vehicle or road side equipment (RSE) can receive packets from another vehicle with a larger-than-zero packet delivery ratio.
Effective Range (ER) is the distance within which the vehicle or RSE can receive packets from other vehicles with a packet delivery ratio larger than a defined threshold.

\subsection{The Simulation of Packet Losses}

The vehicle further away than RMS is more likely to suffer from packet drop during the simulation process where vehicles are all moving on their own tracks. Therefore, we choose to simulate a certain rate of packet drops (lose communication connection). First, we denote the distance from each vehicle member $i$ to the host vehicle as $D_k^i$,, $i \in \{1,2,..., N_v\}$, $ k \in \{1,2,...,t\}$, where $t$ denotes the length of sampling timesteps. Based on whether $D_k^i$ is within ER and MR of the host vehicle, a packet loss rate is assigned according to PDR given above and simulates a situation where the information packets from different vehicles would either be dropped or accepted. Because in the interpenetrating network, we fuse the RBPF mechanism by accepting neighbor information and stack into the host's own particles, the number of packet information transferring is rather large. During this process, the randomness and sudden off-line that may well happen in the real-world vehicle communication is represented.

Regarding maximum range, DSRC radios from different suppliers show some variation, and the average values are around 600 meters with PDR above 10\%. The PDR from data between different months and hours show some difference, but most of the time the PDR are above 70\% for range below 150 meters, which is thus accepted as ER. %However the 10\% percentile results are pretty low, showing that sometimes the DSRC communication is not effective. 
For the vehicles at a distance within MR but outside of ER towards the host, we assign the  PDR in a proportional way. For example, if we used the standard empirical data proposed in \cite{DBLP:journals/corr/HuangZP16}, if at a certain moment, a vehicle is in 375 meters away from the host, then the PDR for this vehicle towards the host would be 40\% at that sampling time.
%  After that, the resampling weights are arranged according to rules and then normalized.

 The pseudo-code is presented as in Algorithm. \ref{alg:pkt_loss}.

\begin{algorithm}
\caption{Packet loss and weight arrangement}\label{alg:pkt_loss}
\begin{algorithmic}[1]
\Procedure{DSRC Packet Loss}{Distance}
% \State Assign a label $ r(i,k)$  to $i$th vehicle at timestep $k$, $ r(i,k) \in [0,1]$;
\State Set a accept probability $ p_{D} = PDR$ according to distance
%, i.e, $p_{CR} = 0.7$, $p_{MR} = 0.1$;
% \State Calculate distance from $i$th vehicle to the host vehicle at timestep $k$, record as $D_k^i$
% \State Calculate particle weights and update $X_t$
\State At timestep k, check if $D_k^i \le$ MR
\While{True}
% check if $r(i,k) \le p_{D}$
% \While{True}
Accept the packet with rate $p_{D}$
%and stack into particle
\EndWhile
\While{False}
Drop this packet
\EndWhile
\State Fuse the packet information into particle
\State Normalize the sum of particle weights to 1;
\EndProcedure
\end{algorithmic}
\end{algorithm}

% (i) At each time step k, assign label as a random number $ lb(i,k) \in [0,1]$ to all packets from the i-th neighboring vehicles, $i \in \{1,2,..., Nv\}$, $ k \in \{1,2,...,Ns\} $;

% (ii) Set a block probability $ p $, i.e, p = 0.3;

% (iii) Check if the distance from host to vehicle I is within confidence range.
% If true, use the packet and go to RBPF;
% If false, check if 'label(i,Ns) $\ge$ block prob': if true, accept the packet and go to RBPF process;
% if false, drop this packet.

% (iv) Check if particles are from “close” neighbors in resample process:
% if true, give higher resample weights;
% if false, give lower resample weights;

% (v) Normalize the distribution and make sure that the weights for one set of particles add up to 1;

\section{DYNAMIC VEHICLE NETWORK FORMULATIONS}
% Combine the dynamic scenario
While most of these studies are focused on mathematical deductions to prove the theoretical feasibility, few works so far have tried implementing this method on a group of vehicles with complete real-life dynamic trajectories, or combine the selection of optimal network configuration and DSRC packet losses existing in real-world traffic scenarios. Here we provide the model with dynamic as well as DSRC packet drops and implemented on an optimized vehicle network structure based on the real-world road condition database.

\subsection{Implementing Position Data for CMM}

\begin{figure}[ht]
\includegraphics[width=\linewidth]{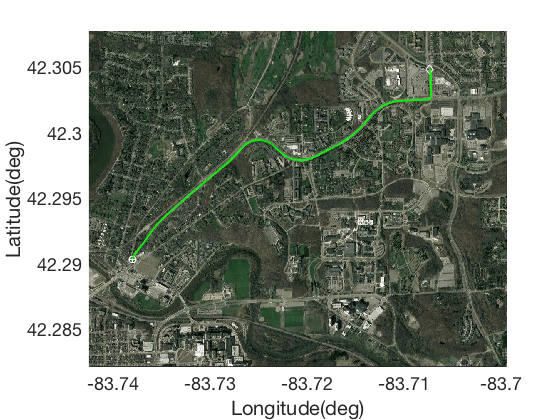}
%24_veh_1st_traj_2.png}
     \caption{Trajectory of the host vehicle on the Google Map. The green line shows how the vehicle travels on the road. The dot with cross shows the position where the simulation starts,and the diamond shows where the simulation ends.}
     \label{fig:host_traj_gglem}
 \end{figure}

The whole trajectory for the host vehicle presented is plotted on Google map for illustration in Fig. \ref{fig:host_traj_gglem}. The green line shows how the vehicle travels on the road. The dot with cross shows the position where the simulation starts, and the diamond shows where the simulation ends.

One thing to be noticed that we treat the traffic data from Safe-Pilot database, such as Latitude-Longitude positions, road angles (i.e., the angle that vehicle trajectories point at) and vehicle velocities to be the true positions and velocities, not "GNSS measured positions/velocities". This is because we do not have any "ground true" data for vehicles on the road to be compared with. Instead, we use these vehicle data to provide a close analog of real-life dynamic scenarios, where vehicles are all traveling on the road at changing locations, speed and directions, and have DSRC communications with each other in real time. A simulated GNSS measurement noise, which is much close to real-life measurement displacement, is alternatively added up to the traveling data using Satellite Navigation TOOLBOX 3.0 to form a real-time GPS measurement of vehicle positions. After that, the RBPF algorithm is applied to get rid of existing common error. 
% As a result, the final results of apparent common error are sometimes biased. It can be possible that the common error contained in Safety-Pilot data itself caused such effects.

\subsection{Simulation Model Formulation}

Here, two basic types of vehicle models (stationary vs. dynamic) are presented, each with multiple simulation scenarios. The effect of Interpenetrating Cooperative Localization method, here the decentralized CMM, is presented with a comparison to traditional centralized CMM. All models use satellite estimation generated with each under Satellite Navigation TOOLBOX
3.0 on Matlab 2017a as the estimated positions for vehicle positions in North Ann Arbor City. The CMM algorithms are then applied with RBPF approach based on each of the proposed configurations of vehicle connections, considering a Dedicated Short-Range Communication (DSRC) of packet-drops varying on vehicle distances. Both models implement real-world driving data from the Safety-Pilot Model Deployment database.
%, which is simulated based on the past three years' driving data of Ann Arbor city, as described in Section IV. A in details.

The first model assumes a stationary situation, where all vehicles are placed at the start positions of the trajectories described in Section V. 
%Basically, this set of performance of vehicle localization is served as a control case to be compared with dynamic results. 
Two issues are presented as control of a single variable respectively, without the influence of the vehicle dynamic on the road: first, the effect of packet drops according to DSRC range; and second, the effect of distributed networks, which used the interpenetrating mechanism, as opposed centralized structure. 
In the simulations, the sampling period and length are taken to be 0.1 s and 3000, respectively. Due to insufficiency of data density,
the max range and effective range of the DSRC are each defined as MR = 4000 meters, ER = 1000 meters.

The second implements real-world driving data with dynamic trajectories. Two types pf simulations are presented: the first simulation assumes dynamic of the host vehicle,
%an ellipse circuit, 
 while all its neighbors are stationary. The second simulation assumes full dynamic network, where all the vehicles conduct trajectories in Fig. \ref{fig:24vehiDistb}, Fig. \ref{fig:38vehiDistb} and Fig. \ref{fig:50vehiDistb}. 
Although it is rear in real life to have all the neighbors stationary except the host, it is worth investigating to see how the dynamic movement of neighbors, can influence the CMM measurement of the host in the real world. Since the effect of packet losses has already be presented, we therefore present only the scenario with Packet drops according to DSRC Range, which better analog the real-world situation for a dynamic host being connected to a network with stationary neighboring vehicles.

% The third model is to implement real-world driving data from the Safety-Pilot Model Deployment database, which is simulated based on the past 3 years' driving data of Ann Arbor city, as described in Section IV. A in details. 

\section{Simulation Results}
 \subsection{Impact of Decentralized Network}
% \subsection{Simple network involving four vehicles}
We first present the simulation results on a simple network involving four vehicles at a cross road to demonstrate that model. Each vehicle can only communicate to itself and one another vehicle. The network connection matrix is as follows:
\begin{equation}
E_C=\left[
\begin{matrix}
1&1&0&0\\
0&1&1&0\\
0&0&1&1\\
1&0&0&1\\
\end{matrix}
\right],
\end{equation}
where $E_C(i,j)=1$ means that vehicle $i$ can receive signals from vehicle $j$.\\
\indent We compare the following two approaches:\\
(1) Weight determined by the variance minimization presented in Section III.C.\\
(2) Constant weight matrix:
\begin{equation}
A=\left[
\begin{matrix}
\alpha&1-\alpha&0&0\\
0&\alpha&1-\alpha&0\\
0&0&\alpha&1-\alpha\\
1-\alpha&0&0&\alpha\\
\end{matrix}
\right],
\end{equation}
where $\alpha=0.5$ would correspond to Eq. (\ref{max_degree}).\\
\begin{table}[ht] 
\caption{Mean Squared Error (MSE) and Variance} % title of Table 
\centering      % used for centering table 
\begin{tabular}{c c c}  % centered columns (4 columns) 
\hline\hline                        %inserts double horizontal lines 
Approach&$\sqrt{Variance}$ (m)&$\sqrt{MSE}$ (m)\\ [0.3ex] 
\hline                    % inserts single horizontal line 
Variance Minimization&0.13&0.58\\ 
Constant $\alpha=0.05$&0.38&0.81\\ [0.3ex]
Constant $\alpha=0.1$&0.33&0.63\\
Constant $\alpha=0.2$&0.28&0.6\\
Constant $\alpha=0.4$&0.21&0.56\\
Constant $\alpha=0.5$&0.2&0.57\\
Constant $\alpha=0.6$&0.21&0.58\\
Constant $\alpha=0.8$&0.33&0.59\\
Constant $\alpha=0.9$&0.46&0.62\\
Constant $\alpha=0.95$&0.68&0.8\\
\hline     %inserts single line 
\end{tabular}
\label{var_err}
\end{table}
Table \ref{var_err} shows the simulation results, which reveals a correlation between the variance and the mean squared error (MSE). In general, as the variance decreases, so does the MSE. Nonetheless, this correlation is not a deterministic relationship. While the variance becomes small enough, the correlation becomes weak. This can be explained by Eq. (\ref{error_decom}). As the variance becomes small, the other term becomes dominant, thus weakening the correlation between the MSE and the variance. Nevertheless, the results provide the possibility to minimize variance as a substitution of error
minimization.

 \subsection{Packet Drop Rate Regrading Distance}

We then consider the complex vehicle networks consist of real-world traffic data as presented in previous sections. Here, Net $N_1$ represent the 50 vehicle network, Net $N_2$ represent the 38 vehicle network, and  Net $N_3$ represent the 24 vehicle network. The distance and packet loss for each network is presented in three different scenarios: stationary, host-only dynamic, and full dynamic from real-world traffic data. 

%The distribution of vehicle distances for all the vehicles to the host, here we use the 1st vehicle in the list, in one of the simulation steps is shown in Fig. \ref{fig:dist_hist}. 

% \begin{figure}[ht]
% \includegraphics[width=\linewidth]{fig/Pkt_drop_dist.png}
% %distance_hist.png}
%      \caption{Distance from all vehicles to the host vehicle in a 50 vehicle network, example given as one of the steps during the simulation process}
%      \label{fig:dist_hist}
%  \end{figure}

It should be noticed, however, we assumed MR  = 4000 meters and CR = 1000 meters in this simulation, with scaling to the results presented in \cite{DBLP:journals/corr/HuangZP16}. This is because we our data used for simulation is not dense enough in the small region within the range  presented in \cite{DBLP:journals/corr/HuangZP16}. It is made sure, however, that the range was extended with a reasonable and equal scaling, which analogs the effect of DSRC packet loss.

The distribution of vehicle numbers within MR and ER of the host together with the packet drop rate precess of the simulation are presented is as in Table. \ref{Packet_Loss}. 
 It should be noticed that the stationary model have a constant number of connected vehicles, while both host-only and full dynamic models are having changing number of neighbors within range, thus receive dynamic connections and packet losses. Please note, however, the packet loss presented is based on the record data given the above-proposed simulation, and is not necessarily the actual packet drop rate in the associated real-life traffics. 
% \textbf{Note: will substitute this with new plots after getting new data}

\begin{table}[ht] 
\caption{Distance and Packet Loss Distribution} % title of Table 
\centering      % used for centering table 
\begin{tabular}{cccccc}  % centered columns (4 columns) 
\hline\hline                        %inserts double horizontal lines 
Networks&Net&Veh. within MR&Veh. within CR&Packet Loss\\ [0.1ex] 
\hline                    % inserts single horizontal line
\multirow{3}{*}{Stationary}& $N_1$ 
&17 &5&34\%\\&$N_2$&13&4&33\%\\&$N_3$&8&2&41\%\\ 
\hline     %inserts single line 
\multirow{3}{*}{Host Dynamic}& $N_1$ &17$ \rightarrow$ 19 $\rightarrow$ 16 &$5\rightarrow 6 \rightarrow 4$ & 37\%\\&$N_2$&13$ \rightarrow$ 10 $\rightarrow$ 12& 4$ \rightarrow$ 2 $\rightarrow$ 3&42\%\\&$N_3$&$8 \rightarrow 10 \rightarrow 9$&$2\rightarrow 4\rightarrow 2$&40\%\\
\hline 
\multirow{3}{*}{Full Dynamic}& $N_1$ &17$ \rightarrow$ 21 $\rightarrow$ 18 &$5\rightarrow 7 \rightarrow 3$ &34\%\\&$N_2$&13$ \rightarrow$ 15 $\rightarrow$ 10&4$ \rightarrow$ 5 $\rightarrow$ 3&37\%\\&$N_3$&$8 \rightarrow 11 \rightarrow 8$&$2\rightarrow 1\rightarrow 3$&39\%\\
\hline 
\end{tabular}
\label{Packet_Loss}
\end{table}

 \subsection{Performance under Stationary Situation}
 
This model assumes a stationary situation, given that every vehicle within the network stays at its starting position of its own trajectory.
The results for two types of simulation models are presented:

% The first one without considering any packet drops during the sampling process, which is an ideal case and consistent with previous works. The second one considered real-time packet losses as described in Section V. A. The third model is the configuration formed using CE that selected 50, 38, 24 vehicles from 1000 candidates, which is used to compare with the first model that use randomly selected vehicle networks directly.

\begin{enumerate}
\item Scenario with zero packet loss. This is theoretically an ideal case without considering any packet drops during the sampling process, which is consistent with previous works on the proposed 24, 38, 50 Network. 

\item Scenario with DSRC packet losses. It considers real-time packet losses as described in Section V, and presented on the proposed 24, 38, 50 Network.

% \item Scenario with CE selected network. The models are formed with configuration selected 24, 38, 50 vehicles from 1000 candidates using CE method, compared with models that use randomly selected vehicles.

\end{enumerate}

\begin{figure}[ht]
\includegraphics[width=\linewidth]{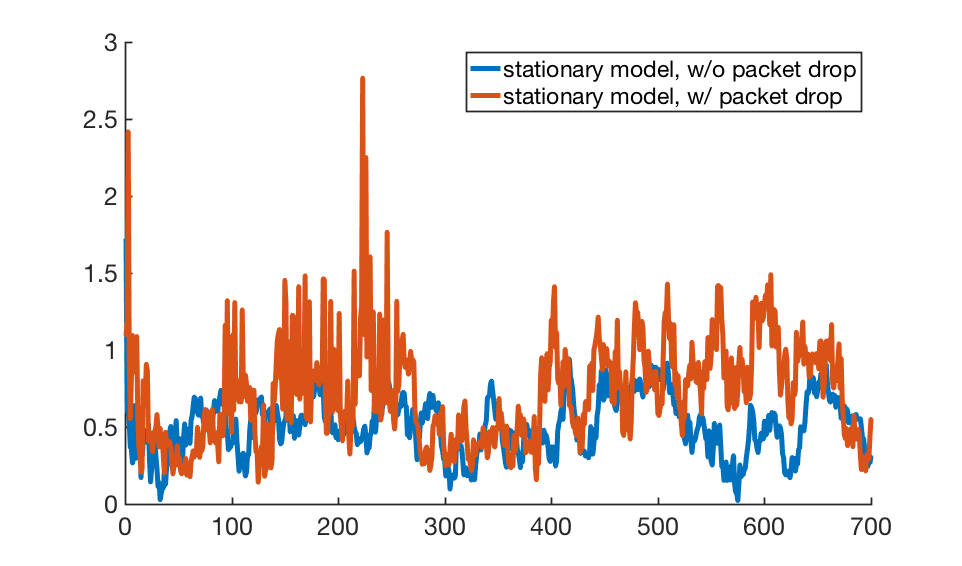}
     \caption{localization error with packet drops (red) vs error without packet drops (blue)}
     \label{fig:pkt_vehi_err}
 \end{figure}

Fig. \ref{fig:pkt_vehi_err} shows the localization error of stationary for a 50 vehicle network over a 700 timesteps sampling period. 
The under two different assumptions: blue line shows the CMM measurement error without packet losses, and the orange line shows the CMM measurement error with packet losses. It can be seen that while the simulation with packet loss is with larger error and more vibrations, the performance is roughly persistent and will not go unbounded.

 \indent The root mean square error (RMSE) of these three CMM mechanisms on the three networks are listed in Table \ref{RMSE_large_net}. 
 
\begin{table}[ht] 
\caption{Root Mean Squared Error (m)} % title of Table 
\centering      % used for centering table 
\begin{tabular}{c c c c c}  % centered columns (4 columns) 
\hline\hline                        %inserts double horizontal lines 
Packet&CMM mechanism&Net 
$\mathcal{N}_1$&Net $\mathcal{N}_2$&Net $\mathcal{N}_3$\\ [0.3ex] 
\hline                    % inserts single horizontal line 
\multirow{3}{*}{W/O Packet Loss}&
Centralized&0.67&0.74&0.86\\ & 
Decentralized Opt.&0.76&1.21&1.44\\&
Decentralized
Rand.&0.72&1.60&4.30\\
\hline
\multirow{2}{*}{W/ Packet Loss}&
Centralized&0.90&1.23&1.53\\&
Decentralized
Opt.&1.44&1.68&2.73\\
&
Decentralized
Rand.&2.03&3.28&4.90\\
\hline     %inserts single line 
\end{tabular}
\label{RMSE_large_net}
\end{table}

 \subsection{Performance under Dynamic Situation}

\begin{figure}[ht]
\includegraphics[width=\linewidth]
{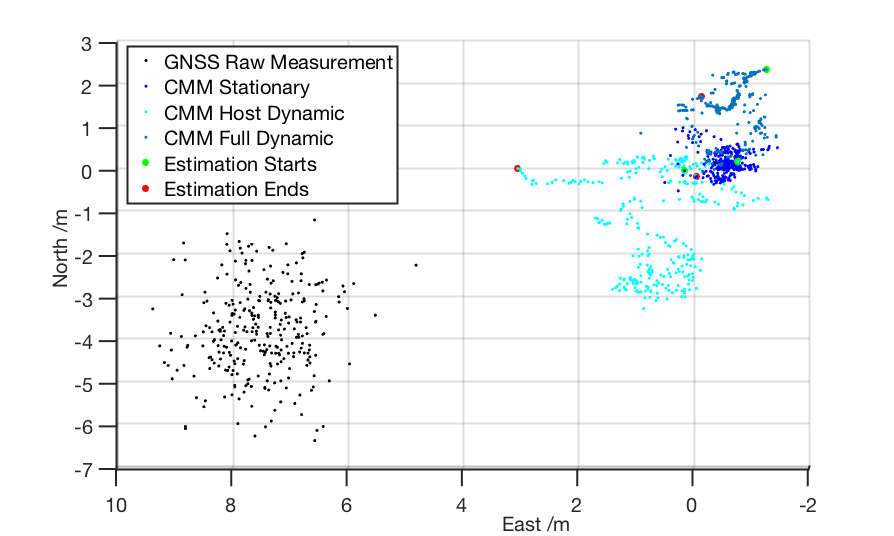}
%{fig/RBPF_RW_24vehi_err_xy.png}
     \caption{Localization error comparison using 24 vehicle network trajectories from Safety Pilot: the black dots are the raw GNSS error, the dark, light, and mild blue dots are stationary, host-only dynamic and full dynamic. We use North-East position as x-y axis, positioning error during whole simulation process of 300 timesteps (5s)}    \label{fig:rw_24_err_vehi_xy}
 \end{figure}

\begin{figure}[ht]
\includegraphics[width=\linewidth]
{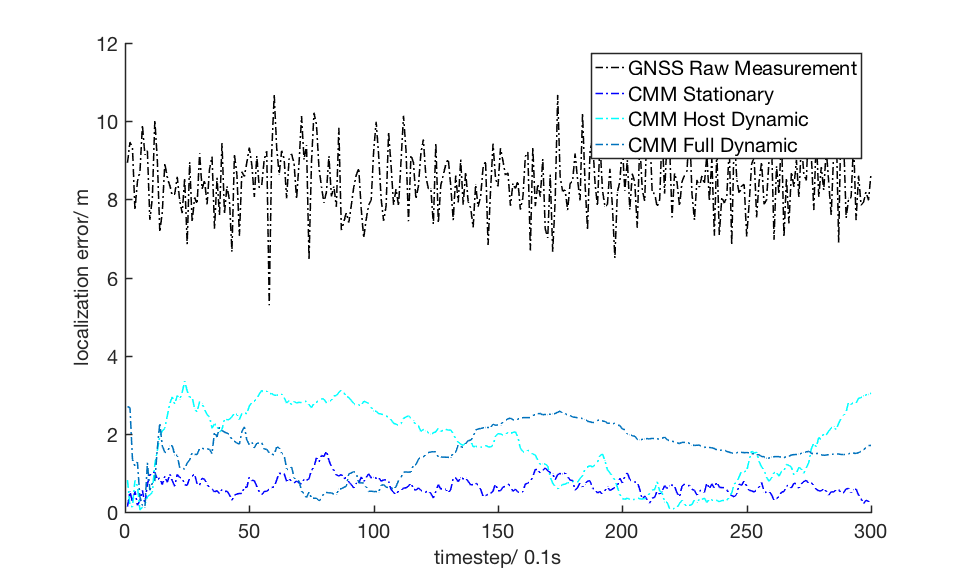}
%{fig/RBPF_RW_24vehi_err_xy.png}
     \caption{Localization error comparison using 24 vehicle network trajectories from Safety Pilot: the black dots are the raw GNSS error, the dark, light, and mild blue dots are stationary, host-only dynamic and full dynamic We use North-East position as x-y axis, positioning error during whole simulation process of 300 timesteps (5s)}    \label{fig:rw_24_err_vehi_t}
 \end{figure}

The dynamic model assumes the vehicles to have its own motions in traffic, each with a different but constant velocity within the legal range, and along their directions predetermined by the associated road-map angles.
% \begin{enumerate}
% \item 
% Performance Host-Only Dynamic Simulation
For the dynamic situation, two simulations are presented.
The first simulation models a host-only dynamic situation, 
% where the host vehicle moves dynamically on the generated trajectory
% and all the neighboring vehicles remain stationary. 
and the second simulation models the fully dynamic vehicle networks according to actual road conditions. Due to time limit, we present the dynamic results with the centralized mechanism.
The position error distribution during the simulation process for the full dynamic network is presented in Fig. \ref{fig:rw_24_err_vehi_xy} and Fig. \ref{fig:rw_24_err_vehi_t}.

The points represent the position errors during sampling time: black points represent raw GNSS measurement, and the blue points are the position errors after applying CMM. Also, the green point is where the RBPF measurement starts (i.e., t = 0), and the red point is where it ends. While the whole sampling period is for 3000 timesteps, we pick the first 300 timesteps to illustrate the error under different situations. Fig. \ref{fig:rw_24_err_vehi_xy} shows the general location displacement of GNSS raw measurement up to a mean of 9 meters, compared with CMM estimated results in stationary and dynamic situations, where the x-axis and y-axis are each for the error placed in North-East directions, and Fig. \ref{fig:rw_24_err_vehi_t} shows the error distribution over the sampling period.

% The experiments are each conducted on two set of networks: vehicle network generated directly with the mechanism described in Section IV.A, and 24, 38, 50 networks selected using CE method from a total 1000 vehicle trajectories. 

The simulation results are presented in Table. \ref{Dyn_large_net}.
 
\begin{table}[ht] 
\caption{Root Mean Squared Error (m)} % title of Table 
\centering      % used for centering table 
\begin{tabular}{c c c c c}  % centered columns (4 columns) 
\hline\hline                        %inserts double horizontal lines 
Packet&CMM mechanism&Net 
$\mathcal{N}_1$&Net $\mathcal{N}_2$&Net $\mathcal{N}_3$\\ [0.3ex] 
\hline                    % inserts single horizontal line 
\multirow{1}{*}{Host Dynamic}&
Centralized&1.89&4.02&3.98
\\
%CE Selection& & &\\
\hline
\multirow{1}{*}{Full Dynamic}&Centralized
&2.03&4.17&3.52\\
%&Decentralized&&&\\

\hline     %inserts single line 
\end{tabular}
\label{Dyn_large_net}
\end{table}

\begin{figure}[ht]
\includegraphics[width=\linewidth]{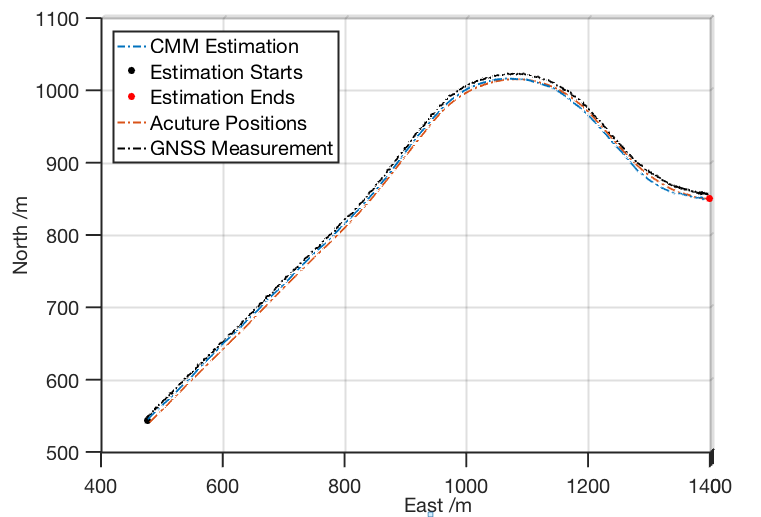}
     \caption{50 vehicle network from Safety Pilot: north-east position as x-y axis; orange line for actual positions, blue line for RBPF filtered positions (Red dot is the start (at 501th timestep), and black dot is the end (at 1500th timestep), black line stands for the direct GNSS estimated position without applying RBPF}
     \label{fig:rw_50_vehi_xy}
 \end{figure}

The trajectories each contains 3000 timesteps (5 min). To show the details of error distribution, we provide the trajectories plot from 501 to 1500 timesteps as an illustration, as shown Figure. \ref{fig:rw_50_vehi_xy}. 
 Here, the orange line stands for actual positions, blue line stands for CMM filtered positions (Red dot is the start (at 501th timestep) and black dot is the end (at 1500th timestep), black line stands for the direct GNSS estimated position without applying RBPF. It can be seen that the CMM estimated trajectory approaches the actual position better than the GNSS measurement even in a complex dynamic case, with the common error being reduced.

\section{Discussion}
From the simulation results, three things are indicated:
First, the distribute CMM using Interpenetrating Cooperative Network provide effective reduction of GNSS measurement errors with less direct connection required, which works both in stationary and dynamic cases. Secondly, the DSRC related packet loss brings more instability to the simulation results, as oppose the ideal case. However, the localization errors do not appear to go unbounded. Thirdly, historical vehicle data abstracted from SPMD database provides good analog to the real-world dynamic traffic situation, which allows us to fulfill the experiment with multiple real-world scenarios for both traditional and interpenetrating CMM. 

One thing to be noticed that we treat the traffic data from Safe-Pilot database, such as Latitude-Longitude positions, road angles (i.e., the angle that vehicle trajectories point at) and vehicle velocities to be the true positions and velocities, not "GNSS measured." An additional GNSS measurement noise is added to the vehicles' data using Satellite Navigation TOOLBOX 3.0.
It remains to be investigated  whether the biases of apparent common error subject to the influence of the original GNSS measurement error contained in Safety-Pilot data. By using these data, however, the simulations in both stationary and dynamic scenarios provide a distinct reduction of GNSS common error, and the simulation errors never go unbounded. 
This indicates that using SPMD data can provide a valid analog of real-life traffic.
% Another thing to be noticed is that there is no distinct difference shown between host-only and full dynamic models. The time delay caused by the sudden off-line and reconnection, however, are not fully simulated in this model. It remains to be investigated, however, whether ...

\section{Conclusion}
This work provides a fusion mechanism for Interpenetrating Cooperative Localization using distributed CMM, and shows the correlation between the estimation variance and the MSE over the network. 
To keep the error small, it is necessary to keep the variance small. Based on this observation, we proposed an optimization approach to determine the fusion weights by solving decentralized quadratic programming. Simulation results verified the correlation between the estimation variance and the MSE. The optimized distributed CMM is shown to have higher accuracy and robustness than a random distributed CMM, especially when the network is sparse. 

Based on that, we implemented complex real-world traffic on both centralized and decentralized CMM. The practical packet losses in the connected vehicle network regarding DSRC is also implemented. The position data are from SPMD database, which imitates the actual traffic scenarios where all vehicles on roads are moving synchronously at changing positions and velocities. These vehicle data are used as an analog to real-life traffic, where vehicles are all traveling on the road at changing locations, speed, and directions, and have DSRC communications with each other in real time. Though the localization displacement is larger than the ideal stationary case, The simulation results on dynamic and complex traffic show effective reduction of common error comparing to GNSS measurement.
$
\\
\\
\\
\\
\\
$

% \appendices
% \section{}
% \section*{Acknowledgment}


\ifCLASSOPTIONcaptionsoff
  \newpage
\fi

\bibliographystyle{IEEEtran}
\bibliography{ref.bib}

\begin{IEEEbiography}
[{\includegraphics[width=1in,height=1.25in,clip,keepaspectratio]{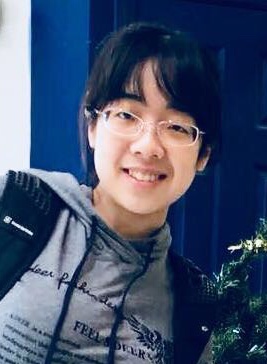}}]{Huajing Zhao} received her M.Eng degree in 2017 from the University of California, Berkeley. She worked in Berkeley Emergent Tensegrity Laboratory as a graduate student researcher during that time. She is currently a M.S. Student at Mechanical Engineering Department of the University of Michigan, Ann Arbor. Her research interest is on Robotics, Automobiles, Artificial Intelligence, and Cognitive Science. 
\end{IEEEbiography}

\begin{IEEEbiography}[{\includegraphics[width=1in,height=1.25in,clip,keepaspectratio]{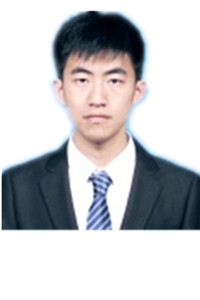}}]{Zhaobin Mo}
received his B. S. degree in 2017 from Tsinghua University, Beijing China. He is currently a research assistant in Tsinghua University, where he is also working toward the Ph.D. in Mechanical Engineering. His research interest includes intelligent transportation, connected vehicle, computer vision and big data analysis.   
\end{IEEEbiography}

\begin{IEEEbiography}[{\includegraphics[width=1in,height=1.25in,clip,keepaspectratio]{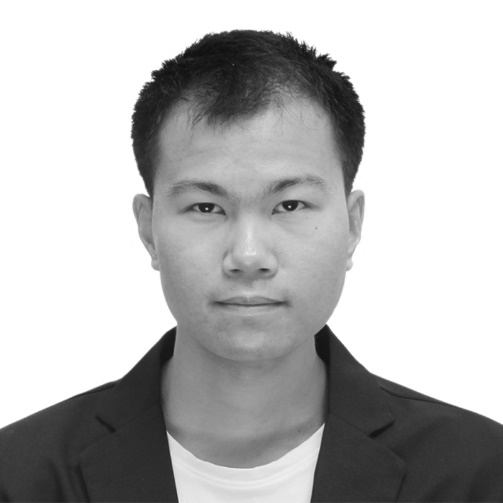}}]{Macheng Shen}
received his B. S. degree in 2015 from Shanghai Jiao Tong University and his M. S. E. degree in 2016 from University of Michigan, Ann Arbor. He is currently a Ph. D. student in Massachusetts Institute of Technology, Cambridge, MA. His research interest includes connected vehicle localization and Bayesian filtering.   
\end{IEEEbiography}

\begin{IEEEbiography}[{\includegraphics[width=1in,height=1.25in,clip,keepaspectratio]{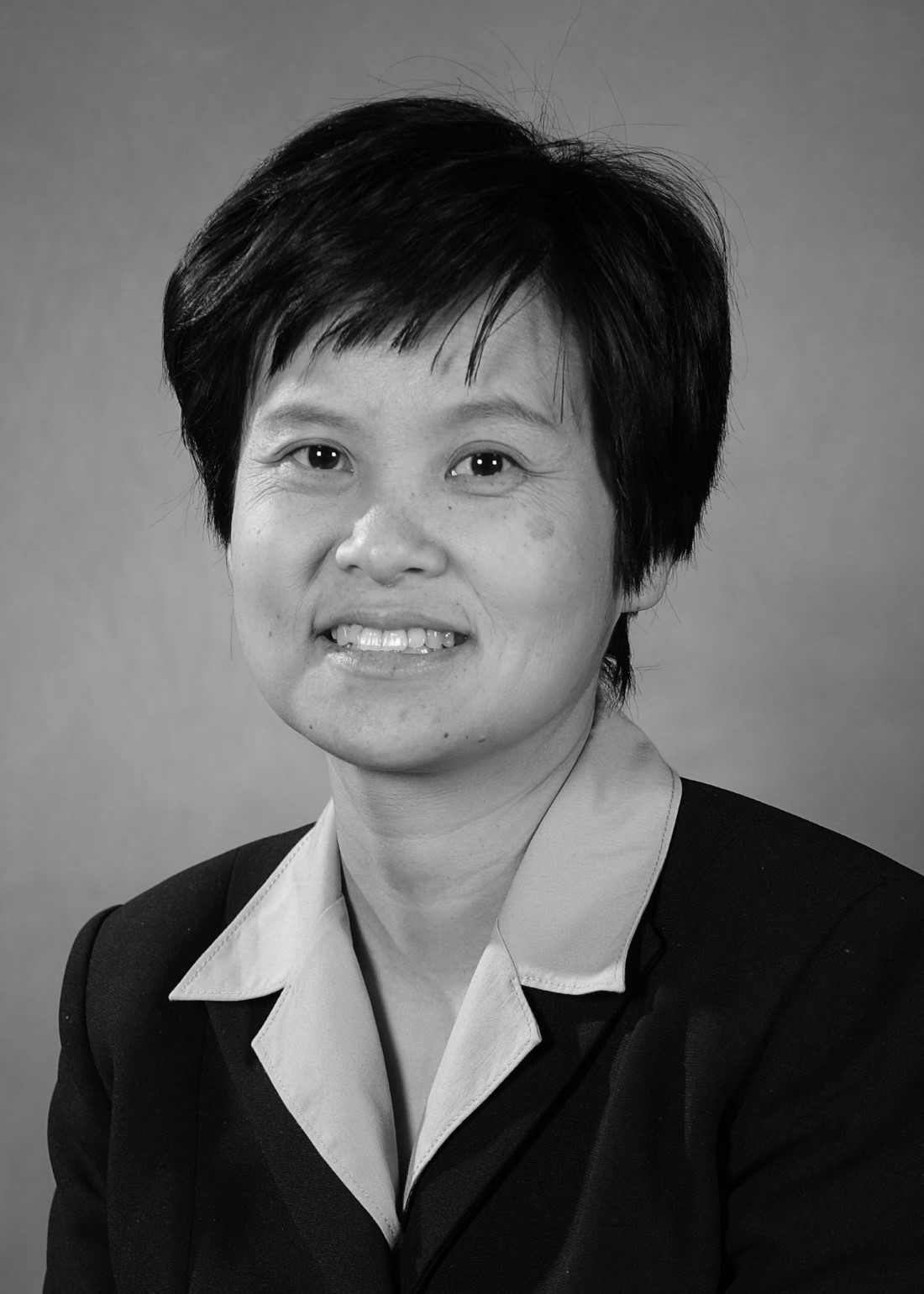}}]{Jing Sun}
received her Ph. D. degree from University of Southern California in 1989, and her B. S. and M. S. degrees from University of Science and Technology of China in 1982 and 1984 respectively. From 1989-1993, she was an assistant professor in Electrical and Computer Engineering Department, Wayne State University. She joined Ford Research Laboratory in 1993 where she worked in the Powertrain Control Systems Department. After spending almost 10 years in industry, she came back to academia and joined the faculty of the College of Engineering at the University of Michigan in 2003, where she is now Micheal G. Parsons Professor in the Department of Naval Architecture and Marine Engineering, with courtesy appointments in the Department of Electrical Engineering and Computer Science and Department of Mechanical Engineering. Her research interests include system and control theory and its applications to marine and automotive propulsion systems. She holds 39 US patents and has co-authored a textbook on Robust Adaptive Control. She is an IEEE Fellow and a recipient of the 2003 IEEE Control System Technology Award.
\end{IEEEbiography}

\begin{IEEEbiography}[{\includegraphics[width=1in,height=1.25in,clip,keepaspectratio]{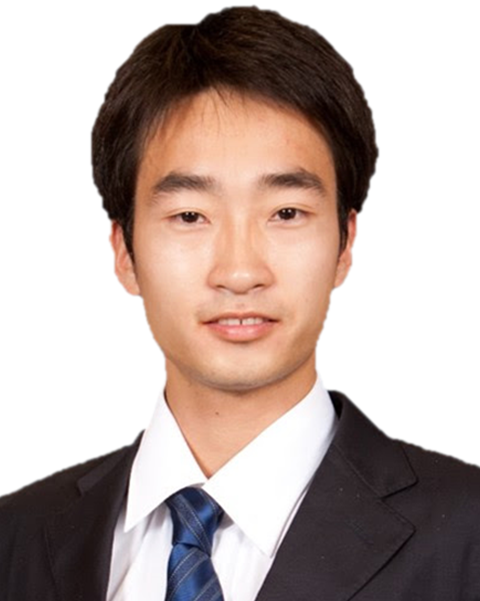}}]{Ding Zhao} received his Ph.D. degree in 2016 from the University of Michigan, Ann Arbor. He is currently an Assistant Research Scientist at Mechanical Engineering of the University of Michigan. His research interests include autonomous vehicles, intelligent/connected transportation, traffic safety, human machine interaction, rare events analysis, dynamics and control, machine learning, and big data analysis.
\end{IEEEbiography}

\end{document}